\documentclass[final,numberedheadings]{aipproc}
\layoutstyle{8x11single}
\usepackage[latin1]{inputenc}
\usepackage[T1]{fontenc}
\usepackage{natbib}
\usepackage{graphicx}
\usepackage{color}
\usepackage{bm}
\usepackage{amsmath}
\usepackage{amssymb}
\usepackage{latexsym}
\usepackage{txfonts}
\usepackage{color}
\newcommand{\be}{\begin{align}}
\newcommand{\ee}{\end{align}}
\newcommand{\bear}{\begin{align}}
\newcommand{\eear}{\end{align}}

\begin{document}

\title{Capillary waves at the interface of two Bose-Einstein condensates.
Long wavelengths asymptotic by trial function approach}

\classification{}
\keywords{}

\author{Todor~M.~Mishonov
}
{address={
Department of Theoretical Physics, Faculty of Physics, 
St. Clement of Ohrid University at Sofia,\\ 
5 J. Bourchier Blvd., BG-1164 Sofia, Bulgaria 
}}

\begin{abstract}
The dispersion relation for capillary waves at the boundary of two different Bose condensates is investigated using a trial wave-function 
approach applied to the Gross-Pitaevskii (GP) 
equations. The surface tension is expressed by the parameters of the GP equations. 
In the long wave-length limit the usual dispersion relation is re-derived while for 
wavelengths comparable to the healing length we predict significant deviations 
from the $\omega\propto k^{3/2}$ law which can be experimentally observed. 
We approximate the wave variables by a frozen order parameter, 
i.e. the wave function is frozen in the superfluid analogous to the magnetic field 
in highly conductive space plasmas. 
\end{abstract}
\date{February 25, 2015}
\maketitle

\section{Introduction}

An interesting property of the Bose Einstein 
condensates is that they can be immiscible \cite{Levin12,Papp08}. 
This distinctive feature opens up the possibility 
to explore and observe interface phenomena. 
The purpose of the present work is to investigate capillary waves at the boundary of two Bose condensates. 
The repulsion interaction between Bose particles is 
modeled by a delta-function Fermi pseudopotential:
\begin{align}
U(\mathbf{r})=g\delta(\mathbf{r}), \qquad g=2\pi\hbar^2a/m,
\end {align}
parameterized~\cite{Sasaki11} by the scattering lengths $a$.
For the system of (1) $^{41}$K and (2) $^{87}$Rb,
for example, we have
\begin{subequations}
\begin{align}
&g_{ij}=2\pi\hbar^2a_{ij}\,(m_i^{-1}+m_j^{-1}),\\
&a_{11}= 65\, a_\mathrm{B},\quad
a_{22}= 99\, a_\mathrm{B},\quad
a_{12}=163\, a_\mathrm{B}.
\end{align}
\end{subequations}
The fluids are immiscible if the xenophilia parameter 
(the inverse relative interaction parameter for bosons)
$\Xi\equiv \sqrt{g_{11}g_{22}}/g_{12}$ satisfies the condition $0<\Xi<1$~\cite{Ao98}, where constants 
$g_{11},\,g_{22},\, g_{12}$ describe the interaction of Bose gases in the 
Gross-Pitaevskii equation.
In order to emphasize the role of the interface in the present work we will analyze the case of the complete segregation $\Xi\ll 1$. 
In the next section~\ref{sec:Gross-Pitaevskii} 
we will recall the Gross-Pitaevskii equation and in section~\ref{sec:Ginzburg-Landau} we will test our Ginzburg-Landau (GL) approach  by reproducing the GL surface tension of a type-I superconductors for the special case of GL parameter $\varkappa\ll1$. 
In section~\ref{sec:Universal} we will analyze the universal Ginzburg-Landau correction to the surface tension which is common for superconductors and Bose condensates.
Later on we will adapt a GL trial function for Bose condensates 
in section~\ref{sec:Trial_Functions}, and following a hydrodynamic analysis in
section~\ref{sec:Hydrodynamics} we will analyze the dispersion relation of the interface 
capillary waves in section~\ref{sec:Dispersion}.

\section{Gross-Pitaevskii equations}
\label{sec:Gross-Pitaevskii}
Our starting points are the Gross-Pitaevskii~\cite{Gross61,Pitaevskii}
equations for the order parameters of two Bose condensates
\begin{subequations}
\begin{align}
\mathrm{i}\hbar \frac{\partial}{\partial t}\Psi_1
&=\left(-\frac{\hbar^2\nabla^2}{2m_1}+V_1+g_{11}\left|\Psi_1\right|^2
+g_{12}\left|\Psi_2\right|^2
\right)\Psi_1,\\
\mathrm{i}\hbar \frac{\partial}{\partial t}\Psi_2
&=\left(-\frac{\hbar^2\nabla^2}{2m_2}+V_2+g_{22}\left|\Psi_2\right|^2
+g_{12}\left|\Psi_1\right|^2
\right)\Psi_2,
\end{align}
\label{TimeDependentGross-Pitaevskii}
\end{subequations}
cf. Sec.~30 in~\cite{LL9}.
These equations are actually Euler-Lagrange equations
\begin{align}
\frac{\delta S}{\delta \Psi_1^*}=0,\qquad
\frac{\delta S}{\delta \Psi_2^*}=0,
\end{align}
minimizing the action
\begin{align}
\label{action}
S=\int \int\left( \mathcal{L}_t-\mathcal{G}_\mathrm{tot}\right)
\mathrm{d}^D \mathbf{r} \,\mathrm{d}t.
\end{align}
The first term of the Lagrangian $\mathcal{L}_t$ 
that has a Schr\"odinger time derivative component has no contribution 
to the further energy balance. 
The second term under the integral is related to the Gibbs free energy.
In detail we have
\begin{subequations}
\begin{align}
&\mathcal{L}_t=\sum_{j=1,2}
\left(\mathrm{i}\hbar \Psi_j^*\frac{\partial}{\partial t}\Psi_j
-\mathrm{i}\hbar \Psi_j\frac{\partial}{\partial t}\Psi_j^*\right),\\
&\mathcal{G}_\mathrm{tot}= \mathcal{G}_1 + \mathcal{G}_2 + g_{12}|\Psi_1|^2 |\Psi_2|^2,\\
&\mathcal G_j=\frac{\hbar^2}{2m_j}
|\nabla\Psi_j|^2
+V_j|\Psi_j|^2+\frac12 g_{jj}|\Psi_i|^4,
\end{align}
\end{subequations}
Here $\mathcal{G}_\mathrm{tot}$ is the Ginzburg-Landau (GL) density of the Gibbs free energy. 
In the 9th volume of the Landau-Lifshitz course on theoretical physics \cite{LL9}, the  
physical conditions and the applicability of Ginzburg-Landau, Ginzburg-Pitaevski and Gross-Pitaevskii 
equations are clearly mentioned and juxtaposed. This is given in practically every textbook on
many-particle physics \cite{Fetter71} or contemporary monographs on Bose-Einstein condensation
\cite{Pitaevskii03,Pethick02}.
In order to test a trial function approach, in the next section we rederive the 
GL result for the surface tension which, according to us, is common for extreme type-I superconductors 
and Bose gases with complete segregation.

\section{Ginzburg-Landau theory for surface tension. Test example for the trial function approach}
\label{sec:Ginzburg-Landau}

Let us now confer to the GL theory~\cite{Ginzburg50} of the surface tension of suprfluids; 
for a pedagogical introduction see Sec.~46 of~\cite{LL9}. 
The density of the Gibbs free energy comprises: kinetic energy of the particles
with effective mass $m^*$ and charge $e^*$, temperature dependent potential $a(T)$, 
a self-interacting term, and energy density of the magnetic field
\begin{align}
\label{GinzburgLandauExFonte}
&
\mathcal{G}=\frac{\left|(-\mathrm{i}\hbar\nabla
-e^*\mathbf{A}/c_\mathrm{light})\Psi \right|^2}{2m^*}
+a(T) |\Psi|^2
+\frac12 b|\Psi|^4
+\frac{1}{2\mu_0}
\left(\nabla \times \mathbf{A}-\mathbf{B}_{\mathrm{ext}}\right)^2,\\
\nonumber& a=a_0\tau,\quad \tau=\frac{T-T_\mathrm{c}}{T_\mathrm{c}},\quad
a_0=\frac{\hbar^2}{2m^*\xi_0^2},\quad \xi
=\frac{\xi_0}{\sqrt{|\tau|}},\\
&\nonumber
e^*=2e, \quad \lambda=\frac{\lambda_0}{\sqrt{|\tau|}},\quad
n_0=\frac{a_0}{b},\quad
\frac{1}{\lambda_0^2}=\frac{n_0}{\varepsilon_0 m^*c_\mathrm{light}^2},\\
&\nonumber\varkappa=\frac{\lambda_0}{\xi_0},\quad
P(T)=\frac{B_\mathrm{c}(T)^2}{8\pi}=\frac12 b \overline{n}^2
=\frac12 a\overline{n},\\
&\nonumber
\overline{n}(T)=-\frac{a}{b}= -\tau n_0,\quad T<T_\mathrm{c},\ 
\mbox{    in Gaussian system } \mu_0=4\pi=1/\varepsilon_0,\\
&\nonumber
\sigma_0=A^*P\xi,\quad \varkappa\ll1,\quad |\tau|\ll1,\quad
\psi_0=\sqrt{\overline{n}}\tanh\left(\frac{z}{\sqrt{2}\xi}\right), \quad z>0,\\
&\nonumber
A^*\equiv2^{5/2}/3=\frac{4\sqrt{2}}{3}\approx 1.89\, .
\end{align}
Here, we have introduced the critical temperature $T_c$, 
the reduced temperature $\tau$, the coherence length $\xi$ 
and its value at zero temperature $\xi_0$, the vector potential $\mathbf{A}$, 
the external magnetic field $\mathbf{B}_{\mathrm{ext}}$, 
the magnetic pressure $P(T)$, the critical magnetic field $B_\mathrm{c}(T)$, 
the penetration depth $\lambda$, the GL parameter 
$\varkappa$ which is supposed to be very small, 
the equilibrium density of the condensate $\overline{n}$ and the surface tension $\sigma$. 
The $\tanh$-profile of the order parameter within the superconductor $z>0$ 
is the solution of the GL equation for small $\varkappa$. 
The test for the correct application of the trial function approach is to rederive the 
$A^*\approx 1.89$ factor in the GL result for the surface tension.
Due to the mathematical similarity, the surface tension between two Bose condensates 
is related to the surface tension of the GL theory. 
Let us consider the interface between the normal and superconducting phases 
parallel to the $x$ and $y$ directions. 
In order to calculate the surface tension, defined as an excess energy per unit area, 
we consider a periodic interface modulation along the $x$-direction 
with magnetic field along $y$-direction
\begin{align}\label{modulation}
\zeta= \zeta_0\sin \left(kx\right),\quad \left\langle\zeta^2\right\rangle_x
=\frac12\zeta_0^2, \quad k\zeta_0\ll1.
\end{align}
Here $\left\langle\dots\right\rangle_x$ denotes an average over $x$. 
We suppose that the amplitude of the modulation $\zeta_0$ is much smaller 
than the wave length $\lambda=2\pi/k$. 
For a rectangular unperturbed surface of area $f_0=L_x L_y$ the interface area change is
$f=L_y \int \sqrt{(\mathrm{d}x)^2+(\mathrm{d}\zeta)^2}$
and its relative increase is
\begin{align}
\frac{\Delta f}{f_0}=\frac{f-f_0}{f_0}
\approx\frac14 k^2\zeta_0^2.
\end{align}
Let us now consider the trial wave function which is a rigid displacement
(in the $z$-direction)  of $\phi_0$:
\begin{align}
\label{grad_ampl}
\Psi(\mathbf{r})=\psi_0\left(z-\zeta(z)\right)
\approx \psi_0(z) - \zeta(z) \,\mathrm{d}_z\psi_0(z).
\end{align}
The first task is to calculate the change of the kinetic energy. 
The order parameter is expressed by the amplitude $\mathcal{A}$ 
and the phase $\theta$
\begin{align}
\label{nabla_psi_2}
\Psi=\mathcal{A}\mathrm{e}^{\mathrm{i}\theta},\quad
\mathcal{A}\equiv\sqrt{n},\quad |\nabla\Psi|^2=|\nabla\mathcal{A}|^2
+ \mathcal{A}^2|\nabla \theta|^2.
\end{align}
For the GL problem the wave function is real-valued and therefore:
\begin{align}
\partial_x\Psi\approx - k\zeta_0 \cos(kx)\, \mathrm{d}_z\psi_0(z).
\end{align}
For the interface, averaged along the modulation direction $x$,
we have up to second order in $\zeta_0$:
\begin{align}
\left\langle(\partial_z\Psi)^2+(\partial_x\Psi)^2\right\rangle_x
\approx \left(1+\frac{1}{2} k^2\zeta_0^2\right)\,\left(\mathrm{d}_z\psi_0(z)\right)^2,
\end{align}
and for the change of the free energy density due to the amplitude gradient we obtain
\begin{align}
\label{G_A}
\left\langle \delta\mathcal{G}_\mathcal{A}\right\rangle_x
=\frac12 k^2\zeta_0^2 \frac{\hbar^2}{2m^*}\left(\mathrm{d}_z\psi_0(z)\right)^2.
\end{align}
In the next steps, we substitute the $\tanh$-profile of the GL solution:
\begin{align}
\mathrm{d}_z\psi_0(z)=
\frac{\sqrt{\overline{n}}}{\sqrt 2\xi\cosh^2(z/\!\sqrt 2\xi)},
\end{align}
and using the integral $\int_0^\infty\cosh^{-4}(t)\,\mathrm{d}t=\frac23$ 
we derive for the change of the free energy per unit area
\begin{align}
&\label{alpha_0}
\nonumber
E_{\mathcal{A},0}
\equiv\int \left\langle \delta\mathcal{G}_\mathcal{A}\right\rangle_x \mathrm{d}z
=\frac{\hbar^2k^2\zeta_0^2\overline{n}}{6\sqrt{2}m^*\xi}
=\sigma_0\frac{\Delta f}{f_0}
=\frac14 k^2\zeta_0^2\sigma_0,
\\
&
\sigma_0=\sigma(k=0)
=2\frac{\hbar^2}{2m^*}\int\left(\mathrm{d}_z\psi_0(z)\right)^2\mathrm{d}z
=-2\frac{\hbar^2}{2m^*}\int
\psi_0(z)\,\,\mathrm{d}_z^2\psi_0(z)
\,\mathrm{d}z
=\frac{4\sqrt{2}}{3}P(T)\,\xi(T)
=A^*\xi P\rightarrow A^*(\xi_1+\xi_2)P\,.
\end{align}
This integral formula is exact and gives the $\psi$-dependent part 
for the surface tension for the exact solution of the GL equations; 
the $\tanh$-profile is only an illustration for the case of $\varkappa\ll1.$
This integral is actually the matrix element of the kinetic 
energy for the motion in $z$-direction.
Therefore, the GL surface tension can be evaluated not only 
as the energy of creation of the phase boundary but also as 
the bending energy of the surface. 
The $\xi P$ representation of the surface tension 
$\sigma$ is common for superconductors and Bose condensates. 
For superconductors $\xi(\tau)$ is the temperature dependent coherence 
length and for Bose gases $\xi$ is the healing length at zero temperature. For the case of complete segregation of two adjacent Bose condensates, one can trivially apply by the substitution $\xi(T)\rightarrow \xi_1+\xi_2$ the GL result for surface tension of the extreme type-I superconductors.
In the next section we will consider the leading correction 
which takes into account the finite penetration in the different phases. 

\section{The universal Ginzburg-Landau correction to the surface tension}
\label{sec:Universal}
For superconductors, the penetration depth $\lambda$ describes the asymptotics of 
the penetration of the magnetic field proportional to $\propto \mathrm{e}^{-z/\lambda}$ 
in the bulk of space homogeneous superconducting phase. 
Analogously, for the case of strong segregation one can introduce the penetration depths 
of one condensate in the bulk of another
\begin{align}
 \psi_1\propto \mathrm{e}^{-|z|/\lambda_1},\quad 
 \lambda_1=\xi_2\sqrt{\Xi},\qquad
 \psi_2\propto \mathrm{e}^{-|z|/\lambda_2},\quad 
 \lambda_2=\xi_1\sqrt{\Xi}.
\end{align}
The geometrical mean is symmetric and indices can be omitted
\begin{align}
 \delta_{_\mathrm{GL}}=\sqrt{\xi\lambda}
 =\sqrt{\xi_1\lambda_1}=\sqrt{\xi_2\lambda_2}
 =\Xi^{1/4}\sqrt{\xi_1\xi_2}\rightarrow \sqrt{\xi(T)\lambda(T)},
 \qquad \lambda=\sqrt{\lambda_1\lambda_2}, \qquad
 \xi=\sqrt{\xi_1\xi_2}
 \,.
\end{align}
The generalization for superconductors is trivial. The variables have to be 
substituted with temperature dependent GL ones.
This GL-length $\delta_{_\mathrm{GL}}$ describes the width of the transition
layer between the different phases when order parameters are small. 
Introducing the dimensionless distance 
$\overline{\tau}\equiv 2^{-1/4} z/ \delta_{_\mathrm{GL}}$ 
the equations for the order parameters for evanescent $\varkappa=\sqrt{\Xi}\ll1$
take the universal GL-form \cite{Ginzburg50}
\begin{align}
\label{universal_GL_system}
 &\mathrm{d}_{\overline{\tau}}^2\ \mathcal{X}
 =\mathcal{Y}^2\mathcal{X},\qquad
 \mathrm{d}_{\overline{\tau}}^2\ \mathcal{Y}
 =\mathcal{X}^2\mathcal{Y},\qquad
 \mathcal{X}(-\infty)=\mathcal{Y}(+\infty)=0,\qquad
 \left.\mathrm{d}_{\overline{\tau}}\mathcal{X}\right|_{\overline{\tau}=\infty}=1
 =- \left.\mathrm{d}_{\overline{\tau}}\mathcal{Y}\right|_{\overline{\tau}=-\infty}.
\end{align}
The $\tanh$ wave function has to be continued by the scaled solution of the 
universal phase boundary function $\mathcal{X}(\overline{\tau}).$ 

There is only one Nobel Prize related to surface tension and we have 
to make a historical remark. Analyzing the $\sqrt{\varkappa}$ correction 
to the surface tension Ginzburg and Landau 
derived this universal system of equations without any parameters 
and concluded that it has to be solved just once \cite{Ginzburg50}. 
Due to the symmetry of the system, 
it is reduced to an universal equation \cite{Mishonov88}
\begin{align}
\label{UniversalPhaseBoundaryEqn}
\mathrm{d}_{\overline{\tau}}^2\ \mathcal{X}(\overline{\tau})
=\mathcal{X}^2(-\overline{\tau})\mathcal{X}(\overline{\tau}),\quad 
\mathcal{X}(-\infty)=0, \quad 
\left. \mathrm{d}_{\overline{\tau}} 
\mathcal{X}\right|_{\overline{\tau}=+\infty}=1.
\end{align}
and the solution gives
\begin{align}
 B^*\equiv2^{9/4}\int_{-\infty}^\infty
(1-\mathrm{d}_{\overline{\tau}}\mathcal{X})\,
\mathrm{d}_{\overline{\tau}}\mathcal{X}\,
\mathrm{d}\overline{\tau}\approx 2.06.
\end{align}
The calculation of further digits should be considered as a routine 
exercise on programming 
by Mathematica product, for example.
Considering $\overline{\tau}$ as a fictitious time 
the corresponding mechanical problem is depicted in \cite{Mishonov88},~Fig.~5.
The constant $B^*$ is proportional to the action of this instanton. 
This instanton action is actually represented by 
the coefficient in the universal GL correction 
to the surface tension \cite{Mishonov88}
\begin{align}
 \Delta \sigma_{_\mathrm{GL}}=-B^*P\delta_{_\mathrm{GL}}, \qquad
 \mbox{ for  } \varkappa=\frac{\lambda}{\xi}=\sqrt{\Xi}\ll 1.
\end{align}
This result is common for both superconductors and Bose gases 
and it is one of the main conclusions of the present research.
Landau became a Nobel Prize winner for his achievements for superfluidity,
while Ginzburg received the same prize for superconductivity.

For superconductors, 
the universal phase boundary function $\mathcal{X}$
describes both the vector-potential $A\propto\mathcal{X}(\overline{\tau})$ 
and the order parameter $\psi\propto\mathcal{X}(-\overline{\tau})$ 
while for Bose gases the same function describes the wave functions 
for both gases $\psi_1\propto\mathcal{X}(\overline{\tau})$ 
and $\psi_2\propto\mathcal{X}(-\overline{\tau})$.
We have to emphasize that this is not an analogy but identity:
equal equations -- equal solutions.

For superconductors, the corrected surface tension \cite{Mishonov88} reads as
\begin{align}
 \sigma_0\approx
 A^*\xi P -B^*\sqrt{\varkappa}\, \xi P, 
\end{align}
while for Bose condensates this result can be rewritten as
\begin{align}
\label{InterfaceTensionBoseCondensates}
 \sigma_0\approx A^*(\xi_1+\xi_2)P
-B^*\frac{\sqrt{\xi_1\xi_2}}{K^{1/4}}P,\qquad
 \frac{1}{K}\equiv \Xi=\varkappa^2\,.
\end{align}
Ginzburg \cite{Ginzburg98} has mentioned the calculation of 
the $\sqrt{\varkappa}$ correction of the surface tension
as a difficult problem and finally gave the credit 
to the solution \cite{Mishonov88} of 
GL universal correction to the surface tension \cite{Ginzburg50}. 
The analyzed correction 
gives an acceptable accuracy for the interpretation of the experimental data.
For the $^{85}$Rb and $^{87}$Rb system one can evaluate the xenophilia parameter for the two configurations 
to be $\Xi=1/2.36$ and $\Xi=1/3.01$ \cite{Papp08}. 
After this brief comment on the problem of interface and surface tension 
of superfluids we will introduce
the trial functions of our dynamical problem.

\section{Trial functions}
\label{sec:Trial_Functions}

For the wave function of the two Bose condensates we will follow the same idea, 
namely to factorize the order parameter by $z$-dependent amplitude and time 
and space dependent phase
\begin{align}
\Psi_j(\mathbf{r},t)=\psi_j(z)\,\mathrm{e}^{-\mathrm{i}\mu t/\hbar}
\mathrm{e}^{\mathrm{i}\theta_j}.
\end{align}
For the rigid wave function we will use the static solution of Gross-Pitaevskii equations
\begin{subequations}\label{Gross-Pitaevskii}
\begin{align}
&\left(-\frac{\hbar^2\mathrm{d}_z^2}{2m_1}+V_1+g_{11}
\left|\psi_1\right|^2+g_{12}\left|\psi_2\right|^2-\mu_1\right)\psi_1(z)=0,\\
&\left(-\frac{\hbar^2\mathrm{d}_z^2}{2m_2}+V_2+g_{22}
\left|\psi_2\right|^2+g_{12}\left|\Psi_1\right|^2-\mu_2\right)\psi_2(z)=0.
\end{align}
\end{subequations}
Note that, in order to derive these static GP equations we actually take into 
account the time derivative term $\mathcal{L}_t$ of the Lagrangian. 
We continue by assuming the idealistic case of complete phase segregation, 
equal intra-species scattering lengths and equal masses such that 
$g_{11}=g_{22}=g,$ $m_1=m_2$ and $g_{12}\gg g.$ 
Eqs.~\eqref{Gross-Pitaevskii} then describe the ground state of a flat 
unperturbed interface. 
The remaining physical quantities are the pressure $P$, 
the chemical potential $\mu$, the healing length $\xi$, 
and the sound velocity $c$ that are connected by well known relations
\begin{subequations}
\begin{align}
&P=\frac12 g \overline{n}^2=\frac12\mu \overline{n},\\
&\mu=g\overline{n}=\frac{\hbar^2}{2m\xi^2}
=mc^2=\frac{\partial P}{\partial \overline{n}},\\
&\xi=\sqrt{\frac{\hbar^2}{2mg\overline{n}}}
=\frac{\hbar}{\sqrt{2}mc}.
\end{align}
\end{subequations}
For the extreme case of complete segregation,
supposing nevertheless long wavelength regime $k\delta_{_\mathrm{GL}}$,
the wave functions coincide with the GL 
wave function for extreme type-I superconductors:
\begin{subequations}
\begin{align}
\psi_{0,1}(z>0)&=\sqrt{\overline{n}}\tanh\left(\frac{z}{\sqrt{2}\xi}\right),\qquad \psi_1(z<0)=0,\\
\psi_{0,2}(z<0)&=\sqrt{\overline{n}}\tanh\left(-\frac{z}{\sqrt{2}\xi}\right),\qquad \psi_2(z>0)=0.
\end{align}
\end{subequations}
The first approach to analyse the interface excitations is to 
use a periodic along the $x$-direction which is rigid along the $z$-direction, 
just as we have done for the GL problem, but here a time dependence multiplier 
has to be added, i.e. the first idea is to use a wave function
\begin{align}
 \Psi_j(\mathrm{r},t)=\psi_j(z-\zeta(x,t))
 \mathrm{e}^{-\mathrm{i}\mu t/\hbar}\mathrm{e}^{\mathrm{i}\theta_j}.
\end{align}

Even for the general case of arbitrary wave-functions
$\Psi_{0,j}(x,z,t)$ the interface displacement $\zeta(x,t)$ can be defined as a 
maximum in $z$-direction of the repulsion energy between two Bose condensates
\begin{align}
g_{12}|\Psi_{0,1}(x,\zeta(x,t),t)|^2 |\Psi_{0,2}(x,\zeta(x,t),t)|^2
\propto \mathcal{X}^2(\overline{\tau})\mathcal{X}^2(-\overline{\tau}).
\end{align}
However, there is a necessary generalization which we will analyse 
in the next section devoted to the hydrodynamics of capillary waves.

\section{One quantum hydrodynamics}
\label{sec:Hydrodynamics}
Let us now consider a potential flow for which the velocity field $\mathbf{v}$ 
is presented by the velocity potential $\Phi$
and by the displacement vector $\mathbf{u}$ (also called deformation in the theory of elasticity)
\begin{align}
\mathbf{v}=\nabla \Phi=\partial_t \mathbf{u},\quad
\Phi=\frac{\hbar \theta}{m}.
\end{align}
The velocity potential is an action per unit mass and for superfluids it is proportional 
to the phase of the order parameter. 
The number density, on the other hand, is proportional to the square of the modulus 
of the wave function $n=|\Psi|^2$.
For the velocity potential we have to use harmonic functions with exponential decay 
$\exp(\pm kz)$ in the bulk of the half-space for every gas 
\begin{eqnarray}
\Phi_j=\Phi_{0,j} \cos(kx-\omega t)\,\exp((-)^j kz).
\end{eqnarray}
For the model case of complete segregation we can use 
a trial function for the velocity potential that is common for both superfluids 
\begin{align}
\label{Incompressible_velocity_potential}
\Phi=\Phi_0 \cos(kx-\omega t)\,e^{-k|z|}\,\mathrm{sign}(z),
\qquad \mathrm{div}\,\mathbf{v}=\nabla^2\Phi=0.
\end{align}
In the regime of complete segregation the density of a superfluid in the ``foreig'' half-space is zero while in the ``own'' half-space we have $\tanh^2$-profile. For the first superfluid $j=1$, the density is nonzero for $z>0$ while for the second superfluid $j=2$ the density is nonzero for $z<0$. The velocity normal to the interface is continuous while its tangential component has a jump along the interface. For capillary waves the two superfluids are slipping relative to one another.

For the sake of definitiveness of our further analysis we will consider the half-space defined by $z<0$, cf. Sec.~12 of~\cite{LL6}.
The gradient of the velocity potential yields
\begin{subequations}\label{velocity}
\begin{align}
v_x&=-k\Phi_0\sin(kx-\omega t)\,\mathrm{e}^{kz}
=\partial_t u_x=\omega u_z\,,\\ 
v_z&= k\Phi_0\cos(kx-\omega t)\,\mathrm{e}^{kz}
=\partial_t u_z=-\omega u_x\,,\\
v^2&=v_x^2+v_z^2= k^2\Phi_0^2\,e^{2kz}.
\end{align}
\end{subequations}
The displacement vector has, therefore, the following components:
\begin{subequations}
\begin{align}
&u_x= -\frac{k}{\omega}\Phi_0\cos(kx-\omega t)\,\mathrm{e}^{kz},\\
&u_z= -\frac{k}{\omega}\Phi_0\sin(kx-\omega t)\,\mathrm{e}^{kz}.
\end{align}
\end{subequations}
Far from the surface $|z|\gg\xi$ the density is almost constant and the 
oscillations of the pressure $\delta P=-m\overline{n}v^2/2$ can be evaluated using the 
Bernoulli theorem. The small (quadratic with the amplitude) variation of the pressure 
also creates small and negligible density oscillations 
$\delta n=\delta P/mc^2.$

Analogously to the derivation of GL surface tension Eq.~(\ref{modulation}) 
one can obtain the boundary condition for the amplitude of the interface modulation 
which gives the relation between the amplitudes of the velocity and the potential oscillations
\begin{subequations}\label{Boundary_Condition}
\begin{align}
&u_{z}(x,z=0,t)=\zeta(x,t)=\zeta_0\sin(kx-\omega t),\\
&\omega\zeta_0=v_0=-k\Phi_0.
\end{align}
\end{subequations}
The first equation is a usual boundary condition for the waves at the surface of a fluid, 
cf. \cite{LL6}.
Here we consider as perturbations both the interface displacement and the velocity potential. 
The deformation vector field is a notion of the theory of elasticity.
During such perturbations the order parameter is frozen in the fluid like 
a magnetic field in plasma with high conductivity \cite{LL8}
\begin{align}
\Psi(\mathbf{r},t)
=\psi_{0}(\mathbf{r}-\mathbf{u}(\mathbf{r},t))\,
\exp\left(\mathrm{i}m\Phi(\mathbf{u}(\mathbf{r},t))/\hbar\right),
\end{align}
Here the index $0$ denotes unperturbed static wave function.
The corresponding fluid density is also frozen $n(\mathbf{r},t)
=n_0(\mathbf{r}-\mathbf{u}(\mathbf{r},t)),$ where $n_0(\mathbf{r})
=|\psi_0(\mathbf{r})|^2.$ The flux of the particles $\mathbf{j}
=n\mathbf{v}$ with $\mathbf{v}=\nabla\Phi=\partial_t\mathbf{u}$ 
and $\nabla^2\Phi=0$, however, obeys the conservation law 
$\partial_t n+\nabla\cdot\mathbf{j}=0$. 
The velocity potential $\Phi$ is the action of the fluid per unit mass, 
$m\Phi$ is the action per atom, and $\theta=m\Phi/\hbar$ is the corresponding phase
of the wave function of the atom.
In other words, in our system of notion and notations we use the concept 
of the order parameter frozen in the fluid, as we mentioned it is analogous to
the Alfv\'en theorem for space plasmas \cite{LL8}.
Returning to the problem of two distinct Bose gases we receive the 
final expression for the trial wave functions 
\begin{align}
\label{trial_final}
\Psi_j(\mathbf{r},t)
=\psi_j(z-u_z(x,z,t))\,\mathrm{e}^{-\mathrm{i}\mu t/\hbar}
\mathrm{e}^{\mathrm{i}\theta_j},
\end{align}
i.e. the order parameter has two explicit $z$-dependences which we will 
take into account in our further hydrodynamic analysis considering 
the kinetic ($\nabla\Phi$-dependent) and potential ($\mathbf{u}$-dependent) 
energy in the superfluid. Under the limit of long wavelengths, as used for 
calculation of the GL surface tension, $u_z(x,z,t)\approx \zeta(x,t)$ 
the density profile moves as a rigid object Eq.~\eqref{grad_ampl} 
and the fluid can be considered as incompressible. 
For capillary waves, upon assumption that the group velocity is much smaller 
than the speed of sound, the compressibility effects are negligible. 
One can continue by constituting Eq.~(\ref{trial_final}) in the GP equation 
and obtaining a solution of the linearized system for the real functions 
$u_z(x,z,t)$ and $\theta(x,z,t)$.

\subsection{Kinetic energy of the superfluid}
Our first step is to calculate the kinetic energy per unit area of one semispace superfluid
\begin{align}
E_\theta=\int^{\infty}_{0} \frac{ mn v^2}{2}\mathrm{d}z,
\end{align}
with velocity field Eq.~(\ref{velocity}) calculated by the gradient of the 
velocity potential; all other energies can be considered as potential ones. 
The kinetic energy corresponds to the phase gradient in the term $\nabla\theta$ in 
Eq.~(\ref{nabla_psi_2}).
Let us consider the GL $\tanh$-profile Eq.~(\ref{GinzburgLandauExFonte})
\begin{align}
n_0(z)=\overline{n}\tanh^2\left(\frac{z}{\sqrt{2}\xi}\right)
=\left(1-\frac{1}{\cosh^2(z/\sqrt{2}\xi)}\right)\overline{n}.
\end{align}
After a simple integration of the surface density of the kinetic energy, 
the GL $\tanh$-profile amounts to:
\begin{subequations}
\begin{align}
\label{KineticEnergy}
&
E_\theta=\frac{\left(1-s\,\mathrm L_2(s)\right)m\overline{n}v_0^2}{2k} 
=E_{\theta,0}\,Z_\rho,\quad
s=2^{3/2}k\xi\, ,\\
&
E_{\theta,0}=\frac{m\overline{n}v_0^2}{2k},\quad
Z_\rho(k)=\frac{\int_0^{\infty} e^{-2k|z|}m|\psi(z)|^2 \mathrm{d}z}
            {m \overline{n} \int_0^{\infty} e^{-2k|z|}\mathrm{d}z}=1-sL_2(s) \,.
\end{align}
\end{subequations}
For brevity, we introduce a convenient dimensionless wave-vector 
$s$ and the functions $L_n(s)$ which we will use in our further analysis. 
The dimensionless factor $Z_\rho(k)$ is not a partition function but rather 
describes the wave-vector dependence of the effective density of the superfluid.

While the kinetic energy ($\propto v^2$) stems from the phase gradient of the wave-function, 
the gradient of the amplitude in Eq.~(\ref{nabla_psi_2}) should be attributed 
to the potential energy in the superfluid, determined by the deformation $u_z$ 
and which is considered in the next subsection. We already applied this approach to the GL 
free energy in Sec.~\ref{sec:Ginzburg-Landau}. The next step is to calculate the potential energy.

\subsection{Potential energy of the curved density. Renormalization of surface tension}
In order to calculate the potential energy of the capillary waves, we replace in the GL problem  
Eq.~(\ref{grad_ampl}) the amplitude of surface modulation with the $z$-dependent amplitude 
of the displacement $\zeta(x)\rightarrow u_z(x,z,t).$
For the $z$-dependent amplitude of the wave-function in Eq.~(\ref{grad_ampl}) 
this substitution reads
\begin{align}
\Psi(\mathbf{r})=\psi_0(z-u_z(x,z,t))
\approx \psi_0(z) - u_z(x,z,t) \, \mathrm{d}_z\psi_0(z).
\end{align}
After this replacement, the space density of the GL free energy in Eq.~(\ref{G_A}) 
acquires an exponential multiplier
$\left\langle \delta\mathcal{G}_\mathcal{A}(z>0)\right\rangle_x \rightarrow
e^{-2kz}\left\langle \delta\mathcal{G}_\mathcal{A}(z>0)\right\rangle_x$
which is the only difference from the original GL problem for type-I superconductor surface tension
\begin{align}
\left\langle \delta\mathcal{G}_\mathcal{A}(z>0)\right\rangle_x 
=\frac{\hbar^2k^2\zeta_0^2}{4m^*}(\mathrm{d}_z\psi_0(z))^2
e^{-2k|z|}.
\end{align}
An additional term arises from $\partial_z u_z(z)$ that has an extra $k$ factor. 
However, we will omit it in our long-wavelength analysis.

Formally, the GL result Eq.~(\ref{alpha_0}) for the surface tension $\sigma_0$ 
can be identified as the long wavelength limit of the potential energy of the surface modulation. 
The $z$-dependence of the displacement vector $u_z$ can be considered as the renormalization 
of the surface tension in Eq.~(\ref{alpha_0})  $\sigma_0\rightarrow \sigma=Z_\sigma\sigma_0.$
A simple integration results for the wave-vector dependent renormalization factor to
\begin{align}
Z_\sigma(k)=\frac{\int_0^{\infty} (\mathrm{d}_z \psi_0)^2e^{-2kz}\,\mathrm{d}z}{\int_0^{\infty} 
(\mathrm{d}_z \psi_0)^2\,\mathrm{d}z}=\frac{L_4(s)}{L_4(0)}.
\end{align}
This renormalization factor describes the decrease of the potential energy (surface tension) 
caused by the exponential decrease of the superfluid amplitude far from the interface
\begin{align}
E_{\mathcal{A},0}\rightarrow E_{\mathcal{A}}= Z_\sigma(k) \, E_{\mathcal{A},0}.
\end{align}
Using time-averaged kinetic and potential energy we continue further with the 
calculation of the dispersion relation of the capillary waves.

\section{Dispersion relation}
\label{sec:Dispersion}

Our consideration of the energy of the capillary waves is based on Eq.~(\ref{nabla_psi_2}) 
that is implicitly incorporated in the action functional Eq.~(\ref{action}). 
Analyzing small amplitude waves we actually use linearized equations and a quadratic 
effective Lagrangian. 
The kinetic energy is proportional to the square of the velocity amplitude 
(see Eq.~(\ref{KineticEnergy})), or $E_\mathrm{kin}=E_\theta\propto v_0^2$ 
while the potential energy is proportional to the square of the interface modulation 
(see Eq.~(\ref{alpha_0})), or $E_\mathrm{pot}=E_\mathcal{A}\propto \xi_0^2$. 
For mechanical oscillations with quadratic action the virial theorem gives 
$E_\mathrm{kin}=E_\mathrm{pot}$. The boundary condition Eq.~(\ref{Boundary_Condition}) 
then  determines the dispersion relation
\begin{subequations}
\begin{align}
&
\rho\omega^2=\sigma k^3, \quad \rho=Z_\rho\rho_0,
\quad \rho_0=m\overline{n}, \quad \sigma=\sigma_0 Z_\sigma,\\
&
\rho(k)=2km\int^{\infty}_0 \left(\psi_0(z)\right)^2\mathrm{e}^{-2kz}\mathrm{d}z,\qquad
Z_\rho(k)=1-sL_2(s),\\
& \sigma(k)
=\frac{\hbar^2}{m^*}\int^{\infty}_0\left(\mathrm{d}_z\psi_0(z)\right)^2
\mathrm{e}^{-2kz}\mathrm{d}z,
\qquad  Z_\sigma(k)=\frac{L_4(s)}{L_4(0)},\\
&
Z_\sigma(0)=1=Z_\rho(0),\quad
s=2\sqrt{2}k\xi.
\end{align}
\end{subequations}
Due to our conservative notations the dispersion relation has the same form 
as it has for the usual capillary waves, cf. Sec.~63 of~\cite{LL6}
\begin{align}
\label{dispersion}
\omega(k)=\sqrt{\frac{\sigma k^3}{\rho}}
=\sqrt{\frac{\sigma_0}{\rho_0}}\, k^{3/2}
\sqrt{\frac{Z_\sigma(k)}{Z_\rho(k)}}.
\end{align}
For long wavelengths we arrive at the usual dispersion of the capillary waves.
The last factor in the equation above is the only one that takes into account the wave-vector dependence 
of the frequency renormalization created by the profile of the order parameter. 
This factorization remains the same for any arbitrary distribution of the effective wave 
function close to the interface and not just for the model case of complete segregation. 
One can easily check that at the long-wavelength limit the group velocity is much smaller 
than the sound velocity $v_\mathrm{gr}=\partial\omega/\partial k\ll c$ and therefore, 
the compressibility effects are negligible. 
It is not surprising that we used the velocity potential for an incompressible fluid 
Eq.~(\ref{Incompressible_velocity_potential}) as a trial function. 
Using a trial function approach we have derived the dispersion for relatively small $s$. 
For the opposite case of $k\xi\gg 1$ we have to go beyond the model 
case of complete segregation and for moderate values of $\Xi<1$  we have to to take 
into account the final width of the mutual penetration of Bose gases. When it is small, 
this parameter can be described by the solution 
of the universal phase boundary equation~\cite{Mishonov88},~Eq.~(19) 
which gives the interpolation formula for the surface tension with 
nonanalytic dependence of the GP parameter $g_{12}$.

\section{Discussion and Conclusion}

To compare with the experiment, we have to additionally evaluate the statistical 
renormalization of the surface tension,
cf. Sec.~23 of~\cite{LL9}:
\begin{subequations}
\begin{align}
&\tilde\sigma (q)=\sigma(q)+T\int_q^{\infty}
 \ln\left[\exp\left(\hbar\omega(k)/T\right)-1\right]\,
\frac{2\pi k\mathrm{d}k}{(2\pi)^2}.\\
&\sigma=\sigma_0-0.13\frac{T^{7/3}\rho_0^{2/3}}{\hbar^{4/3}\sigma_0^{2/3}},
\qquad \mbox{for  } T\rightarrow 0, \, .
\end{align}
It will be instructive to compare and relate our trial function with the functions 
$\tilde u$- and $\tilde v$ defined as follows:
\begin{align}
\nonumber
&
\Psi=\left[\psi_0+\tilde u\mathrm{e}^{\mathrm{i}(kx-\omega t)}
+\tilde v^*\mathrm{e}^{-\mathrm{i}(kx-\omega t)}\right]\mathrm{e}^{-\mathrm{i}\mu t/\hbar}.
\end{align}
\end{subequations}
Solutions to $\tilde u$- and $\tilde v$ can be obtained directly from the exact solution 
of the Gross-Pitaevskii equation, cf. again Sec.~30 in~\cite{LL9}. 
We expect that the trial wave function approach will be asymptotically correct for 
small wave vectors while for $s>1$ one could expect significant deviations. 
For nonzero GL parameter $\varkappa$, the profile of the order parameter can be expressed 
by the universal GL equations~\cite{Mishonov88}. Using this universal solution, 
it will be instructive to see how small inter-species penetration 
influences the dispersion for large wave vectors. 
By introducing 
$\tilde u$- and $\tilde v$ variables one can investigate compressibility effects 
and sound velocity dependent corrections to the dispersion. 
This will be the theme for further analysis. 
It is necessary to calculate $g_{jj}$ dependent terms in the expression wave energy.

We suppose that the most informative function for further analysis is the wave-vector 
dependence of the ratio of $\rho_0 \omega^2(k)/\sigma_0k^3$ versus the dimensionless 
wave-vector $s=2\sqrt{2}k\xi$. It is worthwhile to perform direct numerical analysis 
of the GP equations. For this dependence, after power expansion, the trial function approach gives almost linear dependence
\begin{align}
\frac{\rho_0 \omega^2(k)}{\sigma_{0}\,k^3}
&=\frac{Z_\sigma(s)}{Z_\rho(s)}
\mbox{  versus  } s.
\end{align}
for $\partial\omega/\partial k\ll c$. We do believe that this small deviation of the frequency 
from the usual result for the capillary waves can be experimentally observed. At the end, we wish 
to emphasize that for linear waves the calculation of the dispersion is simplified. We have to 
solve a 1-dimensional problem for the distribution of the order parameters and to substitute 
those ground state functions in the integrals for the matrix elements which will give the waves 
dispersion. In order to derive an exact result applicable for short wavelengths 
we have to calculate some extra terms in the derivatives and the matrix elements of the nonlinear 
terms treated as a perturbation. 
In conclusion we wish to emphasize that the $\tanh$-profile is only an illustration. The one dimensional problem for static distribution of the order parameter only gives the ground state of the system. Perturbation theory gives the secular equation for the dispersion. The coefficients in the characteristic equation are different matrix elements of the GP Hamiltonian.
Finally, the dispersion of the waves will be given as a solution of a square equation.

For different Bose gases, the results from the present study can easily be applied 
by making natural generalizations like
$(\rho_1+\rho_2)\omega^2=(\sigma_1+\sigma_2)k^3$. 
The trial function method easily produces an approximate solution. 
Before an arbitrage by an experiment, it is wise to apply reliable, rigorous, 
insightful and exact methods and to calculate how significant are the deviations 
from the simple linear dependence in the plots ($\omega^2/k^3$ versus $k$) and 
($\sigma_0$ versus $\Xi^{1/4}$); the trial function method is asymptotically exact 
for long wavelength waves.

\section*{Acknowledgments}
The author is expressing his gratitude to Joseph Indekeu for the invitation and the creative atmosphere in Leuven during August 2013 when the present work was started, for attracting his attention to the capillary waves on interfaces between BEC gases and on the k-dependence of the surface tension for mixed BEC, for constructing the initial list of authors and for some suggested references.
Credit also goes to Chang-You Lin, Bert~Van Schaeybroeck and Tran Huu Phat for the critical reading of the manuscript and the numerous questions concerning the approach and the results of this research that helped make this presentation more understandable to the broader public. 
The author is particularly grateful to Nguen Van Thu for presenting  $L_n(s)$ by a hypergeometric function.
Last but not least, the author extends his appreciation to Mehran Kardar for the stimulating discussions in August 2013. 
Different stages of this research have been presented at the Francqui Symposium, Leuven, 20 August 2013, at the 10th Conference of the Society of Physicists of Macedonia, Skopje, 25-28 September 2014, and at the Bulgarian National Scientific Conference on Physics, 
Plovdiv, 10-12 October 2014.

\end{document}